# Deep CardioSound-An Ensembled Deep Learning Model for Heart Sound MultiLabelling

Li Guo, Steven Darvenport and Yonghong Peng

Department of Computing and Mathematics, Manchester Metropolitan University, United Kingdom

Email: L.Guo@mmu.ac.uk, Steven.Davenport@stu.mmu.ac.uk, Y.Peng@mmu.ac.uk

**Abstract**— Heart sound diagnosis and classification play an essential role in detecting cardiovascular disorders, especially when the remote diagnosis becomes standard clinical practice. Most of the current work is designed for single category based heard sound classification tasks. To further extend the landscape of the automatic heart sound diagnosis landscape, this work proposes a deep multilabel learning model that can automatically annotate heart sound recordings with labels from different label groups, including murmur's timing, pitch, grading, quality, and shape. Our experiment results show that the proposed method has achieved outstanding performance on the holdout data for the multi-labelling task with sensitivity=0.990, specificity=0.999, F1=0.990 at the segments level, and an overall accuracy=0.969 at the patient's recording level.

## I. INTRODUCTION

Cardiovascular disease (CVD) is one of the most critical non-communicable diseases caused 17.9 million deaths worldwide(W.H. Organization, 2018). The heart sound signal carries early pathological information about cardiovascular disorders. Listening to the heart sound murmurs (cardiac auscultation) has been well recognised as an effective approach to the early diagnosis of CVDs, because of its non-invasiveness and good performance for reflecting the mechanical movements of the heart and the cardiovascular system. Traditionally, the heart sound auscultation relies heavily on a clinician's manual interpretation. As cardiac auscultation may be compromised by sub-optimal sound quality or hearing impairment, it is usually challenging for a human practitioner to interpret the heart sound accurately. There is also inherent subjectiveness in the heart sound interpretation, which may affect the diagnostic accuracy. As reported in the literature, the accuracy of cardio auscultation by cardiologists is only about 80% (Strunic et al., 2007), with a much lower accuracy of 20%-40% if examined by primary care physicians (Lam et al., 2005). Beyond the dilemma of accuracy, with the recent developments of remote diagnosis technologies and equipment (Blass et al., 2013), it is anticipated that remote heart sound diagnosis will become more prevalent in clinical practices in the coming years. Such a change will generate a large volume of recordings needing interpretation, which will heavily increase the workloads of clinical practitioners who are usually already overloaded.

Both dilemmas impose the need for automatic heart sound analysis to improve diagnosis accuracy and reduce the increasing burden put on clinicians. Digitally, a heart sound is a series of physiological signals and is measured as phonocardiography (PCG). As shown in Figure 1, fundamental heart sounds (FHS) can be segmented into four parts, namely, the first heart sound (S1), the systolic period, the second heart sound (S2) and the diastolic period, with each of them reflecting the movements of different heart components.

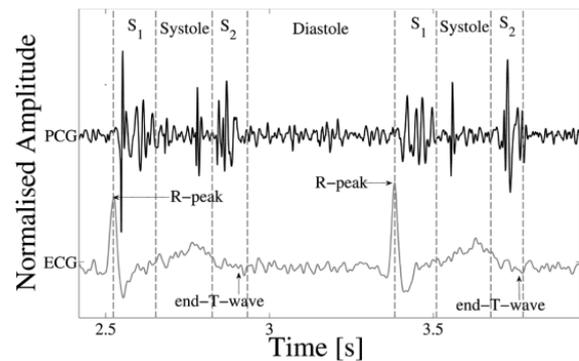

*Figure 1:The four states of the PCG recording: S1, the systole, S2, and the diastole with associated ECG signal(Springer et al., 2016)*

With the PCG, heart sound murmurs (Figure 2) are represented by various waveform patterns that may appear in any PCG segment.

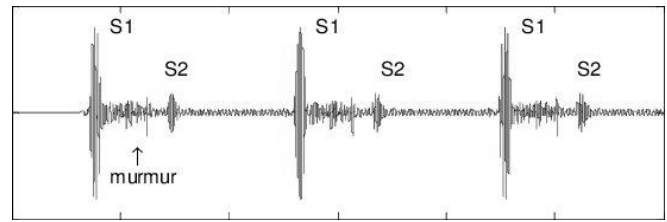

*Figure 2: Example waveform of heart sound murmur*

The typical process for the computer-aided cardiac sound analysis follows typically three steps: waveform segmentation, feature extraction and the development of classification models. For each stage, there have been some existing works reported.

For the segmentation task, (Sun et al., 2014) developed an automatic heart sound signal segmentation method based on the Hilbert transform. In (Springer et al., 2016), the authors proposed to use the hidden semi-Markov model (HSMM) method with logistic regression for segmenting the heart sound signal in a noisy environment. Other proposed methods in recent years include envelop-based methods (Giordano and Knaflitz, 2019; Wei et al., 2019), probabilistic model-based methods (Kamson et al., 2019; Liu et al., 2017; Oliveira et al., 2019) and time/frequency domain based methods(Chen et al., 2017, p. 2; Liu et al., 2018).

After the heard sound signals' segmentation, the segments usually have to go through a feature extraction step that projects the time-domain signals into lower-dimensional feature spaces, facilitating the subsequent classification tasks. A collection of handcrafted feature-based and machine learning-based methods has been intensively studied for this task. Most of the reported work has used Mel frequency cepstrum coefficients (MFCCs) (Abduh et al., 2020; Nogueira et al., 2019), Mel domain filter coefficients (MFSCs), and heart sound spectra (spectrograms) (Soeta and Bito, 2015) that are based on the short-time Fourier transform (STFT), discrete wavelet transform (DWT) coefficients (Deng and Han, 2016) and time & frequency features (Chakir et al., 2018; Potes et al., 2016) from the time-domain, frequency-domain, and time-frequency or scale domain in the S1 and S2 segments.

In the final classification step, fruitful methods have been reported, especially with the deep learning based approaches. The majority of the existing work has used the 1D/2D convolutional neural network (CNN) (Abduh et al., 2020; Alafif et al., 2020; Bozkurt et al., 2018; Chen et al., 2018; Dominguez-Morales et al., 2018, 2018; Humayun et al., 2020; li et al., 2020; Maknickas and Maknickas, 2017; Nilanon et al., 2016; Rubin et al., 2016), recurrent neural network (RNN) (Guo et al., 2019; Khan et al., 2020; Latif et al., 2018; Raza et al., 2019; van der Westhuizen and Lasenby, 2017; Yang and Hsieh, 2016), or the combination of this two (Deng et al., 2020) as their classifiers either with direct use of the raw time domain signals (Humayun et al., 2020; Raza et al., 2019; van der Westhuizen and Lasenby, 2017; Xiao et al., 2020; Yang and Hsieh, 2016)or the feature maps output from the feature extraction stage, including MFCC (Abduh et al., 2020; Bozkurt et al., 2018; Khan et al., 2020; Maknickas and Maknickas, 2017), MFSC (Alafif et al., 2020; Bozkurt et al., 2018) or spectrograms (Dominguez-Morales et al., 2018; Nilanon et al., 2016).

Although many great efforts have been made in this domain, most of the work only focused on the two-class (normal vs abnormal) based heart sound classification task using the PhysioNet/CinC 2016 dataset (Liu et al., 2016). Only a handful number of work has been implemented for multiclass classifications (Baghel et al., 2020; Chen et al., 2018; Chui et al., 2020; Demir et al., 2019; Deperlioglu et al., 2020; Raza et al., 2019) using the 3-classes PASCAL dataset (Bentley et al., 2012) and the 5-classes phonocardiogram Heart sound dataset (Yaseen et al., 2018). However, physiologists often require more information regarding the heart sound in the real clinical environment to better diagnose. According to (Owen and Wong, 2015), the murmuring sound can be classified into several categories based on their occurring locations, grade, pitch levels, qualities and shapes (values of each category are shown in Table 1).

| Timing | Grade | Pitch | Quality | Shape |
|---|---|---|---|---|
| Early Systolic/ Diastolic | I-VI | High | Musical | Crescendo |
| Mid Systolic/ Diastolic | | Medium | Blowing | Decrescendo |
| Late Systolic/ Diastolic | | Low | Harsh | Diamond |
| Holosystolic | | | | Plateau |

Table 1: Different types of murmur labels

Using simplified classification models may work for raising early alarm signals but will be insufficient for more detailed studies. The recent release of the Circor DigiScope Phonocardiogram dataset (Oliveira et al., 2021) has opened new opportunities in this direction.

To further extend the computer-aided heart sound diagnosis domain's landscape, we propose treating the heart sound diagnosis as a multilabel learning problem. Our approach can automatically annotate a heart sound recording with multiple labels, covering various aspects of the cardio auscultation. The main contributions of this work are:

1. To the best of our knowledge, we are the first that uses the multilabel learning method for cardio auscultation.
2. We designed, built, and trained an ensembled deep multilabel learning model with remarkable performance.
3. We conducted a series of experiments to evaluate and validate the model's performance and provided insights into what the model has learned.
4. We performed a deep exploration and analysis of the Circor dataset, which has not been reported anywhere else.

The remainder of this paper is organised as follows. Section II explains our methodology, including problem formulation, network architectures, and design rationales. Experiments and results analysis are presented in section III. In section IV, we presented the performance comparison between our work and the works of literature. We also discuss the challenges and restrictions of the work. Conclusion and future work are discussed in section V.

II. METHODOLOGY

A. *Problem Analysis and Formulation*

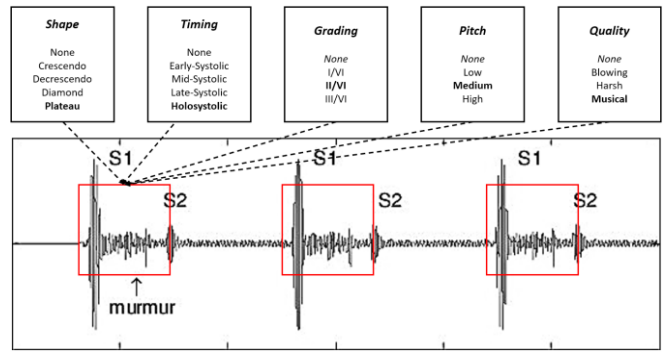

Figure 3: A ensembled multiclass classification problem

Almost all the existing work defines the heart sound diagnosis as a binary or multiclass classification problem, where the heart sound segments/waves are defined as the inputs: $X$ with a list of class labels $y$, where $y_i \in [1, ..., K]$ is the label for the sample $X_i$. With the classifier $f(X; \theta)$, the goal is to find the best $\theta$ values to minimise:

$$argmin_\theta(y - f(X; \theta)) \quad (1)$$

It is important to point out that the values of the class label $y$ are mutually exclusive in this case.

The problem addressed in this work, as shown in Figure 3, is more challenging. We aim to assign multiple labels ($y_{j,i}$) to a heart sound recording while these labels belong to different label groups ($\mathcal{L}_j$). The label values from one label group are mutually exclusive but are non-exclusive to the other labels in the different label groups.

| Timing | | Grading | | Pitch | | Quality | | Shape | |
|---|---|---|---|---|---|---|---|---|---|
| Early-Systolic | 59 | I/VI | 104 | Low | 87 | Harsh | 96 | Plateau | 111 |
| Mid-Systolic | 17 | II/VI | 46 | Medium | 49 | Blowing | 78 | Decrescendo | 34 |
| Late Systolic | 1 | III/VI | 28 | High | 42 | Musical | 4 | Diamond | 31 |
| Holosystolic | 101 | | | | | | | Crescendo | 2 |
| **Total** | **178** | | **178** | | **178** | | **178** | | **178** |

*Table 2: Systolic murmur labelling information of the 178 murmur present patients*

Furthermore, intrinsic relations exist between label groups. For instance, if a recording has a murmur label for the time label group, the labels from the other groups have to be murmur related but cannot be annotated as murmur absent. In contrast, if a recording is labelled as murmur absent for one label group, labels from the rest of the groups must be consistent with it.

This problem can be transformed into a single multiclass classification problem by merging all label groups and creating new classes for each possible combination of the original labels. The overhead with this approach is obvious - an explosion of class numbers ($class_n = i^j$) which is data demanding. We, instead, choose to use an ensemble method that uses a set of classifiers: $y_{j,i} = f_j(X; \theta_j)$ where $y_{j,i} \in [\mathcal{L}_{j,i}, ..., \mathcal{L}_{j,i}]$, with each responding to a single label group. The problem is then formalised as:

$$\sum_j argmin_{\theta_j}(y_{j,i} - f_j(X; \theta_j)) \quad (2)$$

Equation (2), however, only minimises the errors within each label group without considering the interrelations between them. The labels from different label groups may be inconsistent since:

$$argmin_\theta(y - f(X; \theta)) \neq \sum_j argmin_{\theta_j}(y_{j,i} - f_j(X; \theta_j))$$

but:

$$argmin_\theta(y - f(X; \theta)) = argmin_{\theta_j}(y - \omega(f_j(X; \theta_j); w))$$

To address this issue, we introduce another component to the Equation (2):

$$\sum_j argmin_{\theta_j, w}((y_{j,i} - f_j(X; \theta_j)) + (y - \omega(X; \theta_j, w)) \quad (3)$$

The first component in Equation (3) ensures the minimum error achieved for each label group while maintaining the mutual inclusive relations across label groups. The second component controls the global minimum at the concatenated label group level. To approximate the functions $f_j(X; \theta_j)$ and $\omega(X; \theta_j, w)$, we used the convolutional neural networks (CNN) (Lecun et al., 1993) that are illustrated in section II.D.

### B. The Circor DigiScope Phonocardiogram Dataset

The only public dataset with multiple label information annotated is the Circor DigiScope Phonocardiogram dataset (Oliveira et al., 2021). It was collected as part of two mass screening campaigns conducted in Northeast Brazil between July-August 2014 and June-July 2015.

The full dataset contains 5272 heart sound recordings, while only 60% is made public in this paper writing. All the data we used for this study comes from this 60%. This sub-dataset contains heart sound recordings of 942 patients recorded at five different body positions, namely, the pulmonary valve point (PV), the tricuspid valve point (TV), the aortic valve point (AV), the mitral valve point (MV) and the other auscultation location (labelled as 'Phc'). Murmur information is annotated at the patient level instead of the sound segment level. For instance, for a patient that has four recordings at the AV, PV, TV and MV positions, a summarised murmur label set is provided that tells: 1> if murmur appears in the patient recording; 2> the most audible murmur body position; 3> all body positions where murmurs arose (might be multiple positions); 4> murmur's timing; 5> murmur's grading; 6> murmur's shape; 7> murmur's pitch and eight> murmur's quality. Such labels are annotated for the systolic period (from the S1 begin to the systole end) and the diastolic period (from the S2 begin to the diastole end). In addition, for all recordings, segmentation boundaries of the S1 start to the S1 end, the systolic start to the systolic end, the S2 start to the S2 end and the diastolic start to the diastolic end are also provided in the form of timestamps. The sampling rate for all recordings is 4000HZ. The dataset also contains demographic information of the participants, which is not used in this work.

### C. Data Exploration and Data Augmentation

After our initial data exploration, we summarised that out of the 942 patients, 695 patients have no murmur presence, 179 patients have murmur presence, and 68 patients' recordings are marked as 'unknown'. For the non-murmur present patients, their corresponding murmur labels are all recorded as 'nan'. For the murmur presenting patients, only 5 of them have diastolic murmur labels (4 of them also have systolic murmur labels), which are heavily under-presented. Therefore, this study did not consider the S2 and the diastolic periods but focused on the systolic heart sound diagnosis. As a result, the 68 'unknown' patients and the patient (ID: 85119) who only have diastolic labels are excluded from this work. It leaves us with 178 murmur presenting patients. For these 178 patients, the detailed murmur labelling information is summarised in Table 2.

We then assigned murmur label information to the patients' recordings. It should be pointed out that there are multiple recordings from different body positions for each patient, and the murmur signals may or may not be present in all of them. The recordings without murmur presences are filtered out and are treated as normal heard sound signals. The lengths of the recordings vary between 2.13 secs and 66.51 secs, with the mean and median of 22.68 secs and 21.39 secs, respectively. Since we only concentrate on the systolic murmur signals, the segments from the S2 start to the diastolic end are discarded from all recordings.

Furthermore, due to the physiological differences between the patients, the lengths of their S1 start to the systolic end periods are different (max:1.32 secs, min:0.095 secs, mean:0.256 secs, midan:0.259 secs). There is no recording level murmur annotations from the dataset. We cannot precisely capture the locations where the murmurs occur but can only assume that some systolic segments from a recording should contain at least one murmur signal. In other words, the murmur labelling information at the patient level is directly applied to a collection of systolic segments extracted from this patient's recordings.

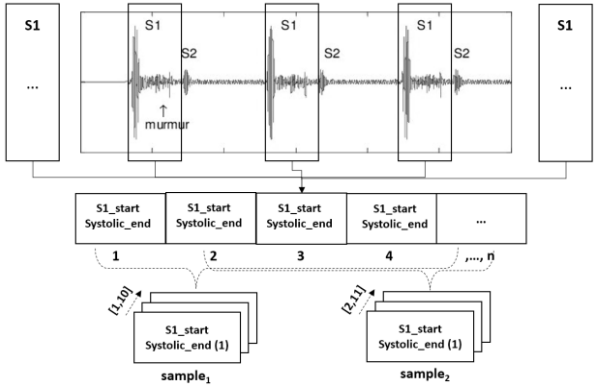

*Figure 4: The data augmentation process*

Another challenge arises from the largely imbalanced samples. The number of recordings with murmur presence is significantly smaller than the non-murmur present ones. *Figure 4* shows our data augmentation method for balancing the samples. We first extract all S1 start to Systolic end segments from a recording and store them in a sequence. Since the median value of the recording is 21.39 seconds and the median value of the S1-Systeloic period is 0.26 seconds, we chose to use 10 segments as the standard size for our data sample. For a recording with no murmur label and containing more than 10 S1-Systolic segments, we randomly draw ten segments from it twice or use all the segments with zero paddings if it has less than 10 S1-Systolic segments.

For a recording labelled as murmur present, we extract all S1-Systolic segments regardless of the recording length. We then applied a sliding window (size=10) over the segment sequence (step=1) to generate multiple data samples. In this way, the number of murmur present samples is boosted. We did not merge all data samples across different body positions since we believe the recordings may contain position-specific properties, which may be hard for a single model to handle (our experiments have verified this).

| Location | Sample No. | Normal-Murmur |
|---|---|---|
| AV | 2147 | 47%-53% |
| PV | 2800 | 39%-61% |
| TV | 2824 | 37%-63% |
| MV | 2812 | 39%-61% |

*Table 3: Data sample size/ratio for each body position*

In addition, the 10 segments in each sample are resampled to 1024 data points with the zero-mean standardisation applied.

### D. The Model Architecture

Figure 5 shows the network architecture that we designed for approximating the function $f_j(X; \theta_j)$. The boxes in lime are data passed into/out from each network layer, and the boxes in blue are network layers.

The network's input is a sequence of 10 S1-Systolic segments, and the network's outputs are multiple one-hot-encoded labels. We use a 121-layer DenseNet (Huang et al., 2016) block as the encoder for feature extraction for each input segment. We choose DenseNet over the other popular network architectures such as ResNet (He et al., 2015)because, with DenseNet, low-level features are pushed to the top of the network and the high-level ones. Also, based on our experience, DenseNet has significantly fewer parameters than the ResNet, making the network more lightweight.

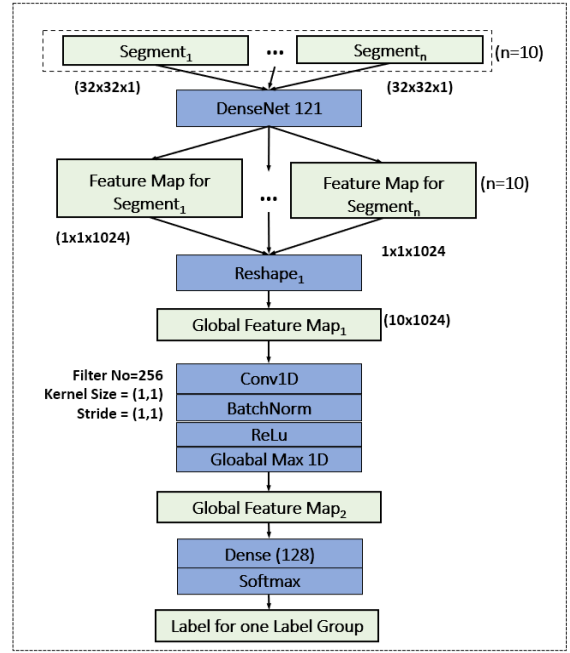

*Figure 5: Network architecture for one label group block*

It is important to point out that the DenseNet block is simultaneously applied to all the input segments. There is only one DenseNet block instance in our network. During the training process, the weights of the DenseNet block get updated using all the gradients backpropagated from its next layer. Unlike a sequence model, no specific order is required for the input segments. This can be understood as follows: the DenseNet block is trained N times in one backpropagation iteration with N data samples.

Similarly, the same filters from the DenseNet block are applied to all segments for feature extraction and generate their feature maps separately during the feedforward phrase. The feature maps of segments are aggregated into a global feature map (Global Feature Map$_1$), which then goes through a 2D convolutional (Conv2D) layer with (1,1) kernel. The conv2d layer combines the filter activations (10x1024) at different scales into the global ones for each receptive view (Global Feature Map$_2$). In this way, the features of all 10 segments are merged for representing the whole input sound wave signals.

For the five label groups (timing, grading, pitch, quality, and shape), we created five identical sub-networks which take the same input and annotate it with the labels from the different label groups (the first component in Equation (3)). There are several considerations with such a network design. First, the features, especially the high-level features required for the labelling task, are unlikely to be similar (Zhang and Wu, 2015).

| fold | AV | | | | TV | | | | PV | | | | MV | | | |
|---|---|---|---|---|---|---|---|---|---|---|---|---|---|---|---|---|
| | pre | sen | spe | F1 | pre | sen | spe | F1 | pre | sen | spe | F1 | pre | sen | spe | F1 |
| 1st | 0.999 | 0.999 | 1.000 | 0.999 | 0.987 | 0.984 | 0.998 | 0.986 | 0.992 | 0.990 | 0.999 | 0.991 | 0.983 | 0.981 | 0.999 | 0.982 |
| 2nd | 0.985 | 0.981 | 0.999 | 0.983 | 0.988 | 0.980 | 1.000 | 0.984 | 0.996 | 0.996 | 1.000 | 0.996 | 0.984 | 0.980 | 0.999 | 0.982 |
| 3rd | 0.992 | 0.984 | 0.999 | 0.988 | 0.999 | 0.999 | 1.000 | 0.999 | 0.995 | 0.995 | 1.000 | 0.995 | 0.995 | 0.992 | 0.999 | 0.994 |
| 4th | 0.996 | 0.993 | 1.000 | 0.994 | 0.994 | 0.989 | 1.000 | 0.991 | 0.992 | 0.989 | 1.000 | 0.991 | 0.993 | 0.991 | 0.999 | 0.992 |
| 5th | 0.984 | 0.984 | 0.998 | 0.984 | 0.996 | 0.995 | 0.999 | 0.996 | 1.000 | 0.999 | 1.000 | 1.000 | 0.992 | 0.991 | 1.000 | 0.991 |
| 6th | 0.992 | 0.990 | 0.999 | 0.991 | 0.995 | 0.993 | 1.000 | 0.994 | 0.991 | 0.988 | 0.998 | 0.989 | 0.981 | 0.976 | 1.000 | 0.978 |
| 7th | 0.984 | 0.982 | 0.998 | 0.983 | 0.996 | 0.994 | 1.000 | 0.995 | 0.992 | 0.990 | 0.999 | 0.991 | 0.987 | 0.984 | 0.998 | 0.986 |
| 8th | 1.000 | 1.000 | 1.000 | 1.000 | 1.000 | 1.000 | 1.000 | 1.000 | 0.996 | 0.996 | 1.000 | 0.996 | 0.992 | 0.991 | 1.000 | 0.992 |
| 9th | 0.986 | 0.984 | 0.998 | 0.985 | 0.993 | 0.992 | 0.999 | 0.993 | 0.998 | 0.997 | 0.999 | 0.998 | 0.990 | 0.988 | 0.999 | 0.989 |
| 10th | 0.991 | 0.988 | 0.999 | 0.989 | 0.994 | 0.989 | 1.000 | 0.991 | 0.996 | 0.996 | 1.000 | 0.996 | 0.983 | 0.980 | 0.998 | 0.982 |
| **Avg** | **0.990** | **0.989** | **0.999** | **0.990** | **0.994** | **0.992** | **1.000** | **0.993** | **0.995** | **0.994** | **0.999** | **0.994** | **0.988** | **0.985** | **0.999** | **0.987** |
| **TD** | **0.993** | **0.990** | **0.999** | **0.991** | **0.995** | **0.991** | **1.000** | **0.993** | **0.994** | **0.994** | **0.999** | **0.994** | **0.989** | **0.987** | **0.999** | **0.988** |

*Table 4: Performance of position-dependent models for the four body positions in the 10 folds*

For instance, the network needs to focus on the murmur location (time-domain) inside a segment for the timing labels. In contrast, for the shape or quality labels, the network needs to learn complicated patterns of the waveforms, and for the pitch labels, the frequency might be the most helpful feature. Expecting a single network to pick up all these different properties would lead to a much more complex network architecture with excessive amounts of parameters and, hence longer training and tuning time. Secondly, as our sub-networks are not interfering with each other (no pathways in between), it gives us engineering flexibility to add new components for future label groups or remove the existing ones without re-training the network. The last benefit of this design is that the sub-networks can be trained in parallel, effectively reducing the training time. The complete network architecture of our work is shown in Figure 6.

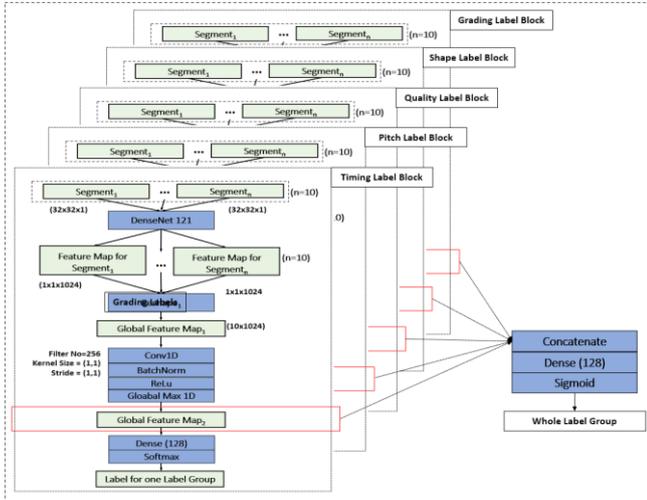

*Figure 6: The complete network architecture*

For the second component of Equation (3)-the function $\omega(X; \theta_j, w)$, we concatenated the global feature map$_2$ from each block, followed by a fully connected layer and used sigmoid as the output function for predicting the whole label group. The network has six outputs, with 5 of them for label groups and the last one for the entire label set. As each subnetwork receives the backpropagated errors from both the label group outputs and the whole label group outputs, the network parameters get updated with a consideration of both. In the feed forwarding phase, however, we only used the outputs from each label block as the final network output.

## III. EXPERIMENTS AND EVALUATIONS

We tested our method through a set of experiments. The model is implemented using the Tensorflow (Abadi et al. 2016) framework (v2.8.0), and the training was carried out on an Nvidia RTX 3080 GPU. We did not apply any feature extraction method (e.g. Mel Spectrogram, MFCC or MFSC) to the input data but only reshaped the raw time-domain signals to a 2D shape, from (10,1024) to (10,32,32,1).

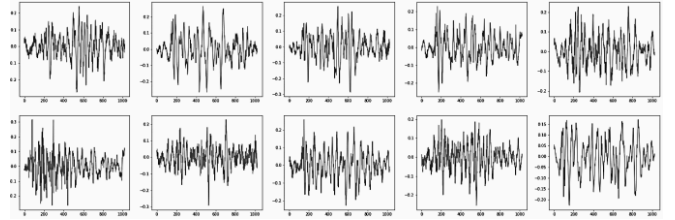

*Figure 7: An input sample from the patient 9947 (PV)*

Figure 7 shows an example input of the network before it is reshaped. For the class labels, we encoded them separately for each label group as timing[0-5], pitch[0-4], quality[0-4], shape[0-5] and grading[0-4] and concatenated them as the final labels. The value 0 in all groups indicates no murmur presence. The Adam optimiser (Kingma and Ba, 2014) is used for training all the models with a learning rate of 0.0001. The batch size used for training is 32, and all the training converged before/at epoch 30. We randomly held out 10% of data (TD) from each body position (AV, PV, TV, MV) for testing and used the remaining 90% of data for training. We followed the standard 10-fold cross-validation process to utilise the training data fully.

### A. Model performance evaluation at the sample level

We first trained 4 position-dependent models, with each for one body position. We used the averaged sensitivity, precision, specificity, and F1 score for all five label groups as the primary metrics to evaluate the model performance at the sample level. Table 4 summarises the overall performances of the models. The results indicate that our network has achieved remarkable performance with the lowest $F1_{avg}=0.987$, $F1_{TD}=0.988$ at the MV position and the highest $F1_{avg}=0.994$, $F1_{td}=0.994$ at the PV position. The averaged metric values across body positions are precision=**0.992**, sensitivity=**0.990**, specificity=**0.999** and **F1=0.991**.

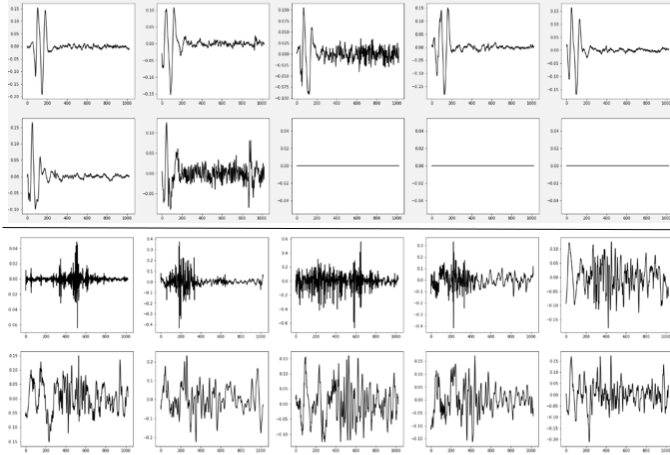 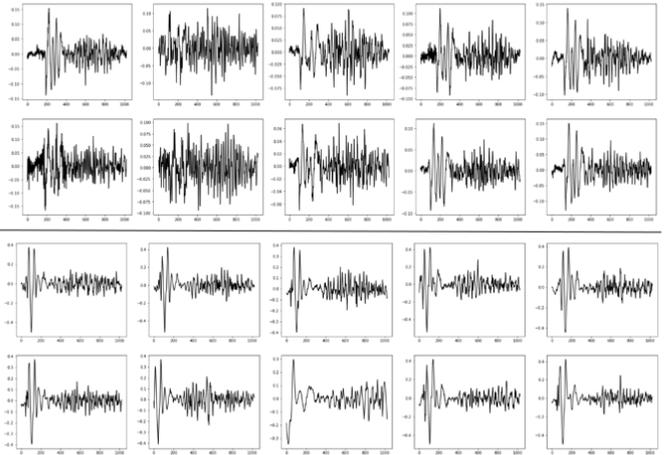

*Figure 8: Mislabelled sample examples from each body position.*
*1) Patient ID: 50238, Location: AV, Timing: **Early Systolic**, Grading: **II/VI**, Pitch: **Medium** Systolic Quality: **Hash**, Shape: **Decrescendo**.*
*Mislabelled as: Timing: Holosystolic, Grading: Normal, Pitch: Normal, Systolic Quality: Normal, Shape: Normal.*
*2) Patient ID: 46065, Location: PV, Timing: **Mid Systolic**, Grading: III/VI, Pitch: Medium Systolic Quality: Hash, Shape: **Decrescendo**.*
*Mislabelled as: Timing: Holosystolic Grading: III/VI Pitch: Medium, Systolic Quality: Hash, Shape: Normal.*
*3) Patient ID: 49823, Location: TV, Timing: **Holosystolic**, Grading: I/VI, Pitch: **Medium**, Systolic Quality: **Hash**, Shape: Plateau.*
*Mislabelled as: Timing: Normal  Grading: I/VI Pitch: Normal Systolic Quality: Normal Shape: Plateau.*
*4) Patient ID:49638, Location: MV, Timing: **Normal**, Grading: Normal, Pitch: Normal Systolic, Quality: Normal, Shape: Normal.*
*Mislabelled as: Timing: Mid-Systolic, Grading: Normal Pitch: Normal Systolic Quality: Normal Shape: Normal*

With the position-dependent models, for the 10,583 samples from all body positions, our network only mislabeled 16. We summarise them in Table 5 with patient IDs. Since these are challenging cases for our models, positioning them may help with future research work from the community. Figure 8 shows 4 mislabelled sample examples. The worst case is the sample of patient 50238 at the AV body position. Our model mislabelled all label groups. However, in our opinion, this is interpretable as this sample contains five normal segments plus three zero-padded segments. The 3$^{rd}$ and 7$^{th}$ segments do not show consistent murmur timing and shape information with the patient level labelling (holosystolic vs early-systolic, plateau vs decrescendo). For the second worst case, the sample of patient 49823, our model failed for 3 label groups (timing, pitch, and systolic quality). It possibly is due to the low frequencies of waveforms in most of the segments (after resample). The model was confused with the timing label and murmur shapes for the other two samples (patients 46065, 49638). For these extreme cases, the network struggled the most with the timing label group, which needs to be further investigated in our future work.

| AV | PV | TV | MV |
|---|---|---|---|
| ID:50238 | ID: 46065 | ID:49823 | ID:49638 |
| ID:84885 | ID: 49968 | ID:50280 | ID:49966 |
|  | ID: 84937 | ID:72283 | ID:50123 |
|  |  | ID:75440 (1) | ID:50331 |
|  |  | ID:75440 (2) | ID:69155 |

*Table 5: Incorrectly labelled samples with their patient IDs*

We also trained a single model using the data from all four body positions (with the same training/testing split) to verify if the model can be body position-independent. The model's performance is reported in Table 6. As expected, the metric results have slightly dropped tough are still better than most of the existing work (please refer to the performance comparison in the discussion section). It indicates that even though the recordings were taken from different body positions, their features do not vary significantly.

In addition, the position-independent model runs four times (avg. 110ms for a 10-second recording) faster at the prediction stage, as it only needs to perform one feedforward. In contrast, the position-dependent ones must make their prediction for each body position in a sequence (avg. 450ms for a 10-second recording). The position-independent model can be used as a lightweight solution for applications with limited resources (e.g. mobile device based).

| fold | precision | sensitivity | specificity | F1 |
|---|---|---|---|---|
| 1$^{st}$ | 0.976 | 0.964 | 0.998 | 0.970 |
| 2$^{nd}$ | 0.973 | 0.968 | 0.997 | 0.970 |
| 3$^{rd}$ | 0.978 | 0.970 | 0.998 | 0.974 |
| 4$^{th}$ | 0.979 | 0.972 | 0.998 | 0.975 |
| 5$^{th}$ | 0.978 | 0.971 | 0.999 | 0.975 |
| 6$^{th}$ | 0.981 | 0.974 | 0.998 | 0.977 |
| 7$^{th}$ | 0.984 | 0.976 | 0.998 | 0.980 |
| 8$^{th}$ | 0.970 | 0.964 | 0.996 | 0.967 |
| 9$^{th}$ | 0.975 | 0.969 | 0.998 | 0.972 |
| 10$^{th}$ | 0.974 | 0.962 | 0.998 | 0.973 |
| **Avg** | **0.977** | **0.969** | **0.998** | **0.973** |
| **TD** | **0.975** | **0.970** | **0.998** | **0.972** |

*Table 6: Performance of the single model for the four body positions in the 10 folds*

### B. Model performance evaluation at the patient level

When we split the sound recordings into segments at the data augmentation stage, we assumed that the segments have the same labels as their containing recordings since there is no segment level labelling information available from the dataset. Although we have used the median of the recordings' lengths for segmentation, it is still anticipated that some segments would be mislabeled, affecting the model performance at the recording/patient level. We conducted an extra evaluation using the models trained at the sample level to label the patient recordings directly.

We first gather all their recordings from different body positions for each patient from the dataset. Only recordings of murmur presence are selected; otherwise, all are chosen. The selected recordings are then segmented by a sliding window (size=10 with no overlap). Based on their body position information, the corresponding network model is applied to the segments and makes its prediction. The sample predictions are finally aggregated using the mode of each label group. The whole process is illustrated in Figure 9.

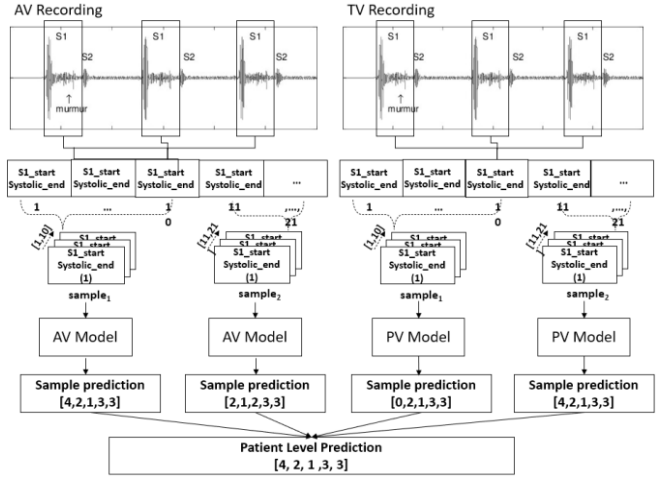

*Figure 9: The process of patient-level prediction*

We used accuracy to measure the difference between the predicted patient labels and the ground truth. That is, given a patient-level label set, $y_{n,j}$ and the predicted label set, $\hat{y}_{n,j}$, the difference between the $y_{n,j}, \hat{y}_{n,j}$ can be measured by:

$$Acc(y_{n,j}, \hat{y}_{n,j}) = 1 - \frac{1}{n} * \sum I(y_{n,j}, \hat{y}_{n,j})$$

$I(y_{n,j}, \hat{y}_{n,j}) = 1$ if $y_{n,j} = \hat{y}_{n,j}$, otherwise 0 if $y_{n,j} \neq \hat{y}_{n,j}$

The final results of the position-dependent models for patient-level labelling are listed in Table 7. Our models scored an overall average accuracy of **96.90%.** For each label group, the average accuracies are 96.96% (timing), 96.89% (pitch), 96.84% (quality), 96.82% (shape) and 96.84% (grading) respectively.

|  | Timing | pitch | quality | shape | grading | Avg |
|---|---|---|---|---|---|---|
| 1st | 96.98 | 96.75 | 96.98 | 96.75 | 96.75 | 96.84 |
| 2nd | 96.98 | 96.75 | 96.98 | 96.98 | 96.98 | 96.94 |
| 3rd | 96.75 | 96.98 | 96.75 | 96.98 | 96.98 | 96.89 |
| 4th | 96.98 | 96.98 | 96.98 | 96.75 | 96.75 | 96.89 |
| 5th | 96.98 | 96.75 | 96.98 | 96.98 | 96.75 | 96.89 |
| 6th | 96.98 | 96.98 | 96.98 | 96.75 | 96.98 | 96.94 |
| 7th | 96.98 | 96.98 | 96.98 | 96.98 | 96.75 | 96.94 |
| 8th | 96.98 | 96.98 | 96.75 | 96.98 | 96.75 | 96.89 |
| 9th | 96.98 | 96.98 | 96.75 | 97.22 | 96.98 | 96.98 |
| 10th | 96.75 | 96.98 | 96.98 | 96.98 | 96.75 | 96.89 |
| Avg | **96.93** | **96.91** | **96.91** | **96.94** | **96.84** | **96.90** |

*Table 7: Performance of the position-dependent models at the patient level*

The performance difference is even more negligible between the position-dependent/independent models at the patient level. From Table 8, we can see that the accuracy of the position-independent model is only 0.42% off its counterparty.

|  | Timing | pitch | quality | shape | grading | Avg |
|---|---|---|---|---|---|---|
| 1st | 96.06 | 96.36 | 96.52 | 96.06 | 96.98 | 96.19 |
| 2nd | 96.75 | 96.75 | 97.22 | 96.06 | 96.75 | 96.71 |
| 3rd | 96.06 | 96.52 | 96.75 | 96.52 | 96.52 | 96.47 |
| 4th | 96.98 | 96.52 | 97.22 | 96.75 | 96.52 | 96.80 |
| 5th | 96.75 | 96.98 | 96.98 | 96.75 | 96.75 | 96.84 |
| 6th | 96.75 | 96.52 | 96.29 | 95.82 | 96.52 | 96.38 |
| 7th | 96.75 | 96.75 | 95.82 | 96.52 | 96.98 | 96.57 |
| 8th | 96.29 | 96.52 | 96.29 | 96.75 | 96.29 | 96.43 |
| 9th | 96.52 | 96.98 | 96.29 | 96.29 | 96.29 | 96.47 |
| 10th | 96.29 | 96.75 | 96.06 | 95.82 | 96.52 | 96.29 |
| Avg | **96.52** | **96.57** | **96.54** | **96.33** | **96.61** | **96.52** |

*Table 8: Performance of the position-independent model at the patient level*

### C. Model Interpretability and Visualisation

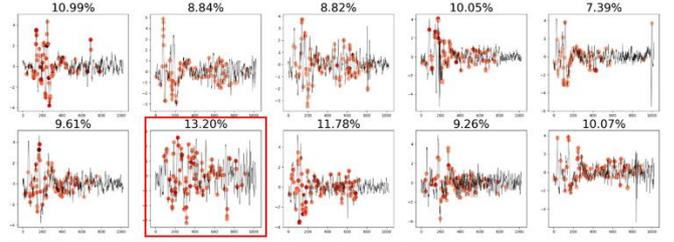

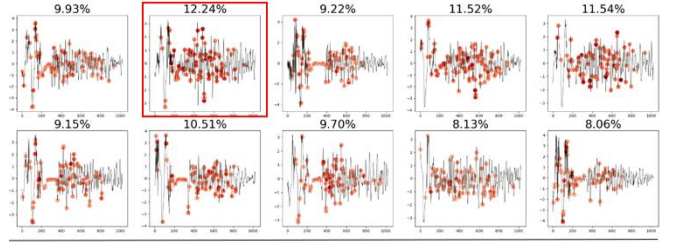

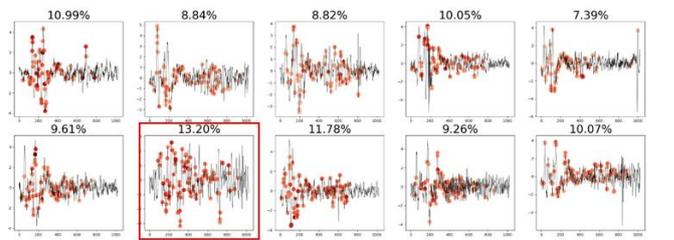

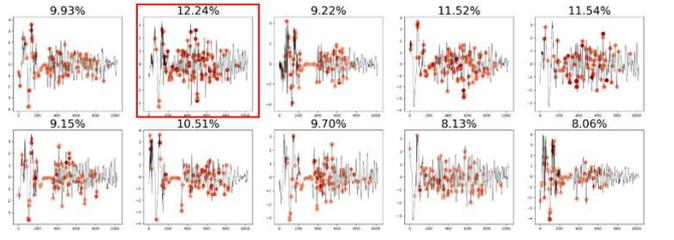

*Figure 10: Example saliency maps of four input samples (with/without murmur presence)*

Even though our model has shown promising results, it is still essential to understand what the model has learned and how it makes its decisions. We used the saliency map (Simonyan et al., 2013) to outline the model's attention areas. Figure 10 shows the attention areas of four sample examples from different patients and body positions with varying labels of murmur. It should be pointed out that, unlike the image-based applications, a single saliency map is not sufficient to explain the attention areas of our model as each model input contains ten segments, and they all contribute to the final labelling results. To handle this, we generated separate saliency maps for the ten segments in a single sample and calculated the contribution of each segment using the sum of its saliency values. With this method, we can visualise the importance of each data point from a segment and show the overall significance of a particular segment in its group to the final labelling.

| Reference | Method | Inputs | Segment Required | No. of Class/Labels | Sensitivity | Specificity | Accuracy |
|---|---|---|---|---|---|---|---|
| (Maknickas et al., 2017) | 2D-CNN | MFSC | No | 2 (N, A) | 0.806 | 0.877 | * |
| (Alafif et al., 2020) | 2D-CNN | MFCC | Unknown | 2 (N, A) | * | * | 0.895 |
| (Deng et al., 2020) | CNN+RNN | MFCC | Yes | 2 (N, A) | 0.987 | 0.980 | * |
| (Abduh et al., 2020) | 2D-CNN | MFSC | No | 2 (N, A) | 0.893 | 0.97 | 0.955 |
| (Rubin et al., 2016) | 2D-CNN | MFCC | Yes | 2 (N, A) | 0.728 | 0.952 | * |
| (Doming et al., 2018) | 2D-CNN | Spectrograms | Yes | 2 (N, A) | 0.932 | 0.951 | 0.970 (selected dataset) |
| (Xiao et al., 2020) | 1D-CNN | Time series | No | 2 (N, A) | 0.853 | 0.957 | 0.936 |
| (Latif et al., 2018) | RNNs | MFCC | Yes | 2 (N, A) | 0.925 | 0.981 | 0.976 (training accuracy) |
| (Li et al., 2019) | 1D-CNN | Spectrograms | No | 2 (N, A) | * | * | 0.965 |
| (Noman et al., 2019) | Ensemble CNN | 1D time-series signals + MFCC | Yes | 2 (N, A) | 0.899 | 0.864 | 0.892 |
| (Chen et al., 2018) | 2D-CNN | Wavelet transform + Hilbert features | No | 3 (N, M, EXT) | 0.980 | 0.885 | 0.93 |
| (Demir et al., 2019) | 2D-CNN | Spectrograms | No | 3 (N, M, EXT) | * | * | 0.800 |
| (Raza et al., 2019) | LSTM | Time series | No | 3 (N, M, EXT) | * | * | 0.808 |
| (Deperlioglu et al., 2020) | AEN | Time series | No | 3 (N, M, EXT) | * | * | 0.960 (best at N) |
| (Chui et al., 2020) | 1D-CNN Wavelet | Time series | No | 5 (N, AS, MS, MR, MVP) | 0.925 | 0.981 | **0.970** |
| (Baghel et al., 2020) | 1D-CNN | Time series | No | 5 (N, AS, MS, MR, MVP) | * | * | 0.962 |
| This work (sample level, position-dependent) | 1D&2D-CNN | Time series | Yes | **22** labels | **0.990** | **0.999** | * |
| This work (sample level position-independent) | 1D&2D-CNN | Time series | Yes | **22** labels | 0.969 | 0.998 | * |
| This work (patient-level, body-dependent models) | 1D&2D-CNN | Time series | Yes | **22** labels | * | * | **0.969** |
| This work (patient-level, body-dependent models) | 1D&2D-CNN | Time series | Yes | **22** labels | * | * | 0.965 |

*Table 9: Performance comparison between our models and the existing works*

For instance, for the sample of patient 14241, the 7$^{th}$ segment (highlighted in red) in the sample made the most contribution (13.20%) to the final result, and the model's attention area was focused on the early systolic period (with most of the saliency points) of the segment. The magnitudes and cluster shape of the saliency points also indicate the pitch level (low) and murmur shapes (plateau). Similar observations are obtained from the samples of patients 49712 and 49628, where the attention areas are highlighted by the saliency points with different clustering shapes (murmur shape: diamond) and magnitudes (murmur pitch: high).

Noticeably, based on the model's region of interest, the 7$^{th}$ segment of the sample of patient 49712 has no murmur presence and only contributes 5.66 % to the classification result, which indicates the model does pick up the correct attention area for making its decision. For the sample of patient 23625 (no murmur presence), the region of interest of the model is mainly around the S1 period, with very few saliency points shown in the systolic period. It is consistent with how cardio physiologists diagnose heart sound recordings in practice.

IV. DISCUSSION

Since there are no standard test benches to which we can directly compare this work, we carried the performance comparison between our models and the existing results focused on the heart sound classification task (single label). Table 9 shows that even though our work deals with many more labels, its performance outperforms those designed for the simple classifications with the best sensitivity=0.990, specificity=0.999 and the top 2 for the accuracy.

That said, there are still a few limitations of this work. First, the performance of our model relies on the accuracy of the upstream segmentation step. It does not handle heart sound recordings directly, whereas some of the existing work for classification tasks can, without requiring this extra step. However, using unsegmented inputs comes with a primary concern- the large number of wrong labels. Many have used a sliding-window based approach to segment the raw signals and have assumed that the segmented raw signals share the same label as the containing recording. Such an approach may work well with heavily murmur presenting recordings but would be suboptimal if murmur presences are low. Although our approach faces a similar issue with the pre-processed segments, the number of wrong labels is significantly lower than that of randomly chunked counterparties.

Secondly, this work has only focused on labelling the systolic murmurs due to the inadequate data samples for the diastolic periods. We do not think this is a significant challenge. As with our network architecture, each label group is separated from the rest. Given more samples, extending this work to support diastolic murmur labelling is straightforward without affecting the existing model's performance for the systolic murmurs.

The most challenging part of this work was how to resolve the inconsistencies between the labelling results across different label groups. Although we have introduced the constraint component in Equation (3) and have adopted the higher-order approach for modelling the intrinsic relations between the label groups, we still found several inconsistent labels in the mislabelled samples. We have tried various methods, including simple and weighted voting mechanisms, learning inter-relations between predicated labels with an additional network, and learning global features across different label groups. The reported result is from the best-performed approach - merging global feature maps from different blocks. When combining it with the voting mechanism at the patient level, the accuracy did increase from 0.969 to 0.972. However, we do not find it a reliable solution, as it heavily depends on the number of non-murmur present patients and the number of label groups. This matter needs to be further studied as the next step of our work.

V. CONCLUSIONS AND FUTURE WORK

This paper presented our work using a deep learning method for multi-labelling heart sound recordings. To the best of our knowledge, this work is the first of its kind in the community. The reported work has extended the landscape of automatic cardiac auscultation from simple heart sound classification to multilabel learning. It will benefit both the machine learning and the clinical community in remote healthcare delivery. With the proposed network architecture, the models have shown outstanding performance for the task, backed by a series of experiments. Beyond that, we have also explained the challenges that we faced with exclusive multilabel learning and have provided deep insights and analysis of the results.

Our next step is to investigate other methods to reduce inconsistencies between the outputs from different label groups. We also plan to apply the model to the diastolic heart sound waves while the remaining 40% of the PhysioNet Circor dataset is released. Furthermore, since the dataset used in this study was collected using a single type of device, this work is device-dependent. We are working on annotating the other existing datasets to develop a more generic model that is device-independent and is scaling better.